\documentclass[a4paper]{article}
\usepackage{interspeech2009,amssymb,amsmath,epsfig}
\usepackage{graphicx}
\ninept
\def\reg{{\rm\ooalign{\hfil
     \raise.07ex\hbox{\scriptsize R}\hfil\crcr\mathhexbox20D}}}

\title{Complex Cepstrum-based Decomposition of Speech for Glottal Source Estimation}

\makeatletter
\def\name#1{\gdef\@name{#1\\}}
\makeatother
\name{{\em Thomas Drugman $^1$, Baris Bozkurt $^2$, Thierry Dutoit $^1$}}

\address{$^1$ TCTS Lab, Facult\'e Polytechnique de Mons, Belgium \\
$^2$ Department of Electrical \& Electronics Engineering, Izmir Institute of Technology, Turkey\\
{\small \tt thomas.drugman@fpms.ac.be}}

\begin{document}
\maketitle

\begin{abstract}

Homomorphic analysis is a well-known method for the separation of non-linearly combined signals. More particularly, the use of complex cepstrum for source-tract deconvolution has been discussed in various articles. However there exists no study which proposes a glottal flow estimation methodology based on cepstrum and reports effective results. In this paper, we show that complex cepstrum can be effectively used for glottal flow estimation by separating the causal and anticausal components of a windowed speech signal as done by the Zeros of the Z-Transform (ZZT) decomposition. Based on exactly the same principles presented for ZZT decomposition, windowing should be applied such that the windowed speech signals exhibit mixed-phase characteristics which conform the speech production model that the anticausal component is mainly due to the glottal flow open phase. The advantage of the complex cepstrum-based approach compared to the ZZT decomposition is its much higher speed.

\end{abstract}


\section{Introduction}\label{sec:intro}

Homomorphic systems have been developed in order to separate non-linearly combined signals \cite{Oppenheim}. As a particular example, the case where inputs are convolved is especially important in speech processing. Separation can then be achieved by a linear homomorphic filtering in the complex cepstrum domain, which presents the property to map convolution into addition. In speech analysis, complex cepstrum is usually employed to deconvolve the speech signal into a periodic pulse train and the vocal system impulse response \cite{Quatieri}, \cite{Verhelst}. Its typical applications then concern pitch detection, vocoding, formant tracking, pole-zero modeling,... but also reach seismic processing or echo detection \cite{Tribolet}.

In parallel, it has been shown \cite{Doval} that speech is a mixed-phase signal where the maximum-phase contribution corresponds to the glottal open phase while the vocal tract component is assumed to be minimum-phase. In \cite{Bozkurt}, we showed that the Zeros of the Z-transform (ZZT) based technique was able to achieve such an anticausality-based decomposition. On the other side, it has been discussed (\cite{Oppenheim},\cite{Quatieri}) that complex cepstrum can be used for source-tract deconvolution although no approach could achieve this robustly. In this paper, we emphasize the role of windowing on the mixed-phase decomposition quality and show how an appropriate window can improve the glottal source estimation. Taking these precautions into account, the new technique is shown to carry out similar results as ZZT, while it is much faster.


The paper is structured as follows. Section \ref{sec:ACdecomp} presents the theoretical framework of anticausality-based decomposition. The Complex Cepstrum (CC) based technique is introduced and similarities with the ZZT-based method are discussed. Both approaches are viewed as two different ways to reach the same goal: separating the maximum and minimum-phase contributions from a signal Z-transform. Section \ref{sec:Synthetic} exhibits our results on synthetic signals. The impact of windowing, which plays a crucial role on the decomposition quality, is described in detail in \ref{ssec:Window}. Section \ref{ssec:Comparison} highlights the fact that CC and ZZT-based techniques have almost exactly the same behaviour while the first one is much faster. Finally Section \ref{sec:Real} confirms the efficiency of our method on real speech signals.

\begin{figure*}[!ht]
  \centering
  \includegraphics[width=0.9\textwidth]{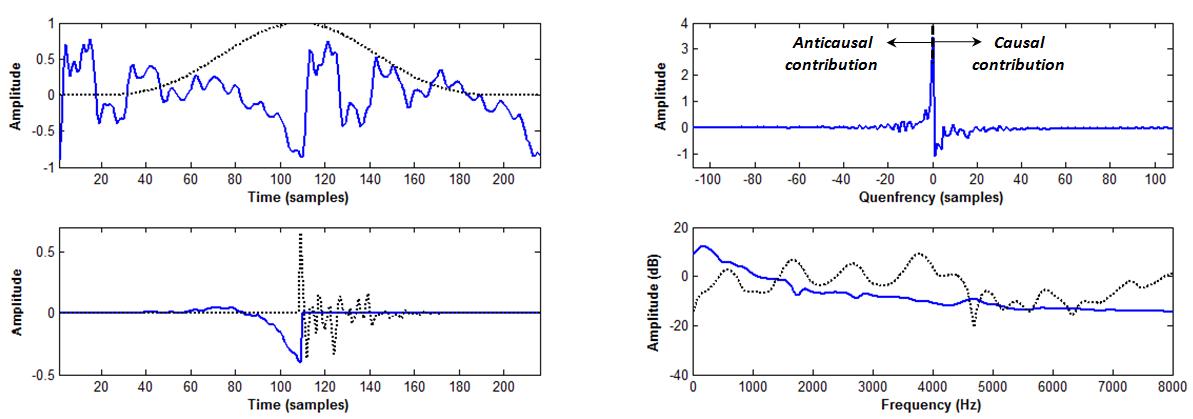}
  \caption{Anticausality-based decomposition of speech using the complex cepstrum. \emph{Top-left}: Real speech signal ($F_s=16kHz$) and applied window. \emph{Top-right}: Corresponding complex cepstrum where the separation of maximum and minimum-phase components can be linearly achieved. \emph{Bottom-left}: Maximum (solid) and minimum-phase (dashed) components. \emph{Bottom-right}: Log-magnitude spectrum of these components.}
  \label{fig:TotalFigure}
\end{figure*}

\section{Anticausality-based decomposition of speech}\label{sec:ACdecomp}

\subsection{Complex Cepstrum based decomposition}\label{ssec:CCdecomp}
The complex cepstrum (CC) $\hat{x}(n)$ of a discrete signal $x(n)$ is defined by the following equations \cite{Oppenheim}:

\begin{equation}\label{eq:DFT}
X(\omega)=\sum_{n=-\infty}^{\infty} x(n)e^{-j\omega n}
\end{equation}

\begin{equation}\label{eq:ComplexLog}
\log[X(\omega)]=\log(|X(\omega)|)+j\angle{X(\omega)}
\end{equation}

\begin{equation}\label{eq:ComplexCepstrum}
\hat{x}(n)=\frac{1}{2\pi}\int_{-\pi}^{\pi}{\log[X(\omega)]e^{j\omega n}}\emph{d}\omega
\end{equation}

where Equations \ref{eq:DFT}, \ref{eq:ComplexLog}, \ref{eq:ComplexCepstrum}  are respectively the Discrete-Time Fourier Transform (DTFT), the complex logarithm and the inverse DTFT (IDTFT). Our decomposition arises from the fact that the complex cepstrum $\hat{x}(n)$ of an anticausal (causal) signal is zero for all $n$ positive (negative). Retaining only the negative part of the CC should then estimate the glottal contribution. An example of separation using the complex cepstrum on a real speech segment is exhibited in Figure \ref{fig:TotalFigure}.

One difficulty when computing the CC lies in the estimation of $\angle{X(\omega)}$, which requires an efficient phase unwrapping algorithm. In this work, we computed the FFT on a sufficiently large number of points (typically 4096) such that:

\begin{itemize}
\item the grid on the unit circle is sufficiently fine, which facilitates the phase evaluation,
\item distortion from aliasing in $\hat{x}(n)$ is minimized. 
\end{itemize}

Besides these phase unwrapping problems, we show in this paper (Section \ref{sec:Synthetic}) that windowing plays a crucial role in the mixed-phase decomposition. 

\subsection{Zeros of the Z-transform based decomposition}\label{ssec:ZZTdecomp}
For a series of $N$ samples $(x(0),x(1),...x(N-1))$ taken from a discrete signal $x(n)$, the $ZZT$ representation is defined as the set of roots (zeros) $(Z_1,Z_2,...Z_{N-1})$ of the corresponding Z-Transform $X(z)$:

\begin{equation}\label{eq:ZZT}
X(z)=\sum_{n=0}^{N-1} x(n)z^{-n}=x(0)z^{-N+1}\prod_{m=1}^{N-1} (z-Z_m)
\end{equation}

Decomposition here arises from the fact that the roots of an anticausal (causal) signal all lie outside (inside) the unit circle. The ZZT-based technique \cite{Bozkurt} aims thus at splitting the roots of $X(z)$ into two subsets $Z_{AC}$ and $Z_{C}$ where the roots have a modulus greater (respectively lower) than one:

\begin{equation}\label{eq:ZZT2}
X(z)=x(0)z^{-N+1}\prod_{k=1}^{M_o} (z-Z_{AC,k}) \prod_{k=1}^{M_i} (z-Z_{C,k})
\end{equation}

\subsection{Unification of both approaches}\label{ssec:Unification}

A parallel can be drawn between both previous techniques since they perform the same operation: separate the minimum and maximum phase contributions from a discrete signal $x(n)$. If $X(z)$ is expressed as in Equation \ref{eq:ZZT2}, it can be shown that \cite{Steiglitz}:

\begin{equation}
\hat{x}(n)= \left\{
\begin{array}{rl}
\sum_{k=1}^{M_o}{\frac{{Z_{AC,k}}^n}{n}} & \emph{if } n < 0\\
\sum_{k=1}^{M_i}{\frac{{Z_{C,k}}^n}{n}} & \emph{if } n > 0
\end{array} \right.
\end{equation}

which confirms the close link between ZZT and CC-based decompositions. Furthermore the phase unwrapping problem can be solved by factorization, as indicated in \cite{Steiglitz2}.

\begin{figure*}[!ht]
  \centering
  \includegraphics[width=0.9\textwidth]{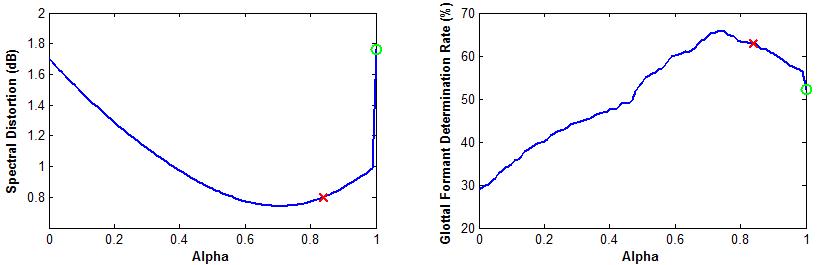}
  \caption{Sensitivity of (\emph{Left panel:}) the spectral distortion and (\emph{Right panel:}) the glottal formant determination rate to the window function. Blackman (x) and Hanning (o) windows are indicated.}
  \label{fig:AlphaonBoth}
\end{figure*}

\section{Tests on synthetic speech}\label{sec:Synthetic}

This Section presents decomposition results on synthetic speech signals for different test conditions. The idea is to cover the diversity of configurations one could find in natural speech by varying all parameters over their whole range. Synthetic speech is produced according to the source-filter model by passing a known train of Liljencrants-Fant glottal waves \cite{Fant} through an auto-regressive filter extracted by LPC analysis on real sustained vowel uttered by a male speaker. As the mean pitch during these utterances was about 100 Hz, it reasonable to consider that the fundamental frequency should not exceed 60 and 180 Hz in continuous speech. Table \ref{tab:Range} summarizes all test conditions.

\begin{table}[!ht]
\centering
\begin{tabular}{c | c}
\hline
Pitch & 60:20:180 Hz \\
\hline
Open quotient & 0.4:0.05:0.9 \\
\hline
Asymmetry coefficient & 0.6:0.05:0.9 \\
\hline
Vowel & /a/, /@/, /i/, /y/ \\
\hline
\end{tabular}
\caption{Table of synthesis parameter variation range.}
\label{tab:Range}
\end{table}


Decomposition quality is assessed through two objective measures:

\begin{itemize}

\item {\bf Spectral distortion} : Many frequency-domain measures for quantifying the distance between two speech frames $x$ and $y$ arise from
the speech coding litterature. Ideally the subjective ear sensitivity should be formalised by incorporating psychoacoustic effects such
as masking or isophone curves. A simple relevant measure is the spectral distortion (SD) defined as:

\begin{equation}\label{eq:SD1}
SD(x,y) = \sqrt{\int_{-\pi}^\pi(20\log_{10}|\frac{X(\omega)}{Y(\omega)}|)^2\frac{\emph{d}\omega}{2\pi}}
\end{equation}

where $X(\omega)$ and $Y(\omega)$ denote both signals spectra in normalized angular frequency. 

\item {\bf Glottal formant determination rate} : The amplitude spectrum for a voiced source (such as the LF model)
generally presents a resonance called \emph{glottal formant}. As this parameter is an essential feature of the glottal open phase, an error
on its determination after decomposition should be penalized. For this, we define the \emph{glottal formant determination rate} as the proportion of frames for which the relative error on the glottal formant frequency is lower than 10\%.

\end{itemize}

This formal experimental protocol allows us to reliably assess our techniques and to test their sensivity to different factors influencing the decomposition.

\subsection{The influence of the window}\label{ssec:Window}

Windowing is known to be a critical issue for obtaining an accurate complex cepstrum \cite{Quatieri}, \cite{Verhelst}. Most approaches discuss the validity of the convolutional model in order to separate the periodic pulse train and the vocal system response. The goal of this Section is to give a complete empirical study of the influence of the window parameters on our particular task: the anticausality-based decomposition. 

\subsubsection{The window position}\label{sssec:WindowPosition}

In \cite{Quatieri} the need of aligning the window center with the system response is highlighted. Although we discuss this issue in \cite{Drugman}, analysis in this work is performed on windows centered on the Glottal Closure Instant (GCI), as this particular event demarcates the boundary between the causal and anticausal responses, and the linear phase contribution is removed.

\subsubsection{The window function}\label{sssec:WindowFunction}

To the best of our knowledge, the impact of the window function on the complex cepstrum was never analyzed. However we showed in \cite{Bozkurt2} that the window has a considerable effect on the root location in the Z-plane and that usual windows (such as Hanning or Hamming) are  generally not the best-suited. In this work, we consider windows of $N$ points satisfying the form \cite{Oppenheim}:

\begin{equation}\label{eq:win}
w(n)=\frac{\alpha}{2}-\frac{1}{2}\cos(\frac{2\pi n}{N-1})+\frac{1-\alpha}{2}\cos(\frac{4\pi n}{N-1})
\end{equation}

for which the Hanning and Blackman windows are particular cases (for $\alpha=1$ and $\alpha=0.84$ respectively). Figure \ref{fig:AlphaonBoth} exhibits the influence of parameter $\alpha$, regulating the window function, on the decomposition performance. In both graphs, an optimum clearly emerges for $\alpha=0.72$. This value is used throughout the rest of the paper. It can be noticed that the widely-used Hanning window is not appropriated for our application.



\subsubsection{The window length}\label{sssec:WindowLength}
\begin{figure}[!ht]
  \centering
  \includegraphics[width=0.4\textwidth]{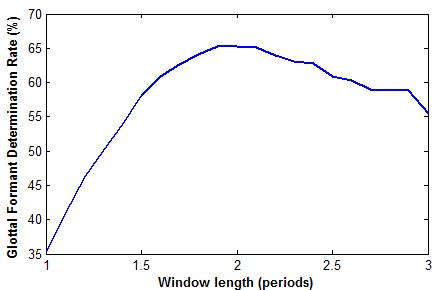}
  \caption{Sensitivity of the glottal formant determination rate to the window length.}
  \label{fig:LengthOnFg}
\end{figure}

In \cite{Quatieri} and \cite{Verhelst}, it is argued that a window whose duration is about 2 to 3 pitch periods gives a good trade-off for being consistent with the convolutional model. Figure \ref{fig:LengthOnFg} shows the impact of the window length on the glottal formant determination. The best decomposition is achieved for two period-long windows. Besides a slight error on the pitch estimation can be tolerated since this would not have a dramatic incidence.

\begin{figure*}[!ht]
  \centering
  \includegraphics[width=0.9\textwidth]{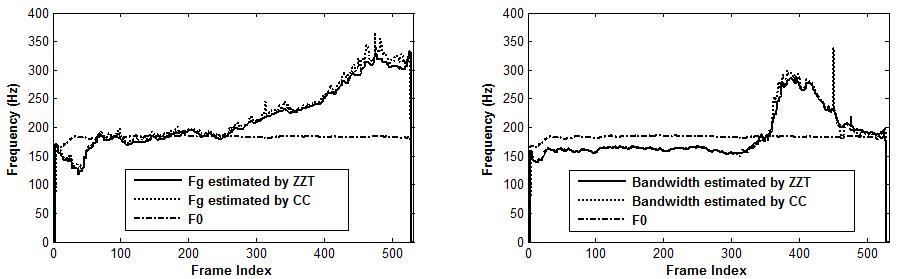}
  \caption{Glottal formant characteristics estimated by both ZZT and CC-based techniques on a real sustained vowel. \emph{Left panel:} evolution of the glottal formant frequency, \emph{Right panel:} evolution of the glottal formant bandwidth.}
  \label{fig:RealSpeech}
\end{figure*}

\subsection{Comparison between CC and ZZT-based decomposition}\label{ssec:Comparison}

As underlined in Section \ref{ssec:Unification}, CC and ZZT-based techniques can be viewed as two different means to separate the minimum- and maximum-phase components from the speech signal. Their decomposition should then be theoretically strictly equivalent. We confirm it through our experiments, except for some rare cases for which decomposition may differ. One possible explanation is the difficulty in reliably unwrapping the phase for these cases. In the following points, we discuss the factors affecting the efficiency of these methods, and compare them in terms of computational load. 

\subsubsection{The factors influencing the decomposition}\label{sssec:Pitch}

Many factors may affect the quality of our decomposition. Intuitively, one can think about the \emph{interference} between minimum and maximum-phase contributions. The stronger this interference, the more difficult the decomposition. Basically, this interference is conditioned by three parameters:

\begin{itemize}
\item the pitch $F_0$, which governs the spacing between two successive vocal system responses,
\item the first formant $F_1$, which influences the minimum-phase time response,
\item and the glottal formant $F_g$, which controls the maximum-phase time response.
\end{itemize}

A strong interference then implies a high pitch, with low $F_1$ and $F_g$ values. Throughout our experiments we confirmed the performance degradation with the evolution of these factors, equally touching CC and ZZT-based techniques. Results yielded by both methods over the whole test set were sensibly identical. A minor difference was reported in favor of ZZT probably for the reasons mentioned above.


\subsubsection{Computational considerations}\label{sssec:Computational}

Since we are treating speech frames whose length is twice the pitch period, the ZZT-based technique requires to compute the roots of generally high-order polynomials (depending on the sampling rate and on the pitch). Although current polynomial factoring algorithms are accurate, the computational load remains problematic. On the other hand, the CC-based method just relies on FFT and IFFT which can be fastly computed. Table \ref{tab:F0OnTime} shows the clear advantage to use the complex cepstrum. 

\begin{table}[!ht]
\centering
\begin{tabular}{c | c | c}
 & ZZT-based & CC-based \\
  & decomposition & decompostion \\
\hline
60 Hz & 1837.1 & 17.11 \\
\hline
180 Hz & 184.7 & 16.49 \\
\hline
\end{tabular}
\caption{Comparison of the required computation time (in $ms$, for our Matlab implementation with $F_s=16kHz$) for different pitch values.}
\label{tab:F0OnTime}
\end{table}


\section{Tests on real speech}\label{sec:Real}

We validate our method on the same example as in \cite{Bozkurt3}. It consists of a sustained vowel \emph{/a/} with a flat pitch and
decreasing open quotient (voluntarily produced by an incresing pressed effort). Figure \ref{fig:RealSpeech} compares the glottal formant characteristics (frequency and bandwidth) estimated by both ZZT and CC-based methods. It can be noticed that here again decompositions give sensibly similar results.

\section{Conclusion}
This paper discussed the methodology for applying the complex cepstrum for mixed-phase decomposition, allowing the estimation of the glottal source. The importance of a suited windowing has been highlighted. It has been shown how the applied window conditions the separation quality. A parallel with the ZZT-based technique has been drawn since both methods aim at separating minimum and maximum-phase contributions of speech. Results we obtained for both methods are sensibly similar, while the complex cepstrum-based decomposition is much faster.

\section{Acknowledgments}\label{sec:Acknowledgments}

Thomas Drugman is supported by the ``Fonds National de la Recherche
Scientifique'' (FNRS).

\eightpt
\bibliographystyle{IEEEtran}

\end{document}